\documentclass[pra,a4paper,twocolumn,superscriptaddress,showkeys]{revtex4}

\usepackage{latexsym}
\usepackage{amssymb}
\usepackage{amsmath}
\usepackage{enumerate}
\usepackage{bm}
\usepackage{paralist}
\usepackage[latin1]{inputenc}
\usepackage{color,xcolor}
\usepackage{graphicx}
\graphicspath{{./figs/}}
\usepackage{subfigure}
\usepackage[normalem]{ulem}

\newcommand {\queq}[1]{(\ref{#1})}
\newcommand {\qeq}[1]{Eq.~\queq{#1}}
\newcommand {\qeqs}[1]{Eqs.~\queq{#1}}
\newcommand {\qfig}[1]{Fig.~\ref{#1}}

\newcommand {\qtable}[1]{Table~\ref{#1}}

\newcommand {\ns}{n_{\textrm{surf}}}

\newcommand {\etal}{\emph{et al.}\ }

\newcommand {\beql}[1]{\begin{equation} \label{#1}}
\newcommand {\eeql}{\end{equation}}
\newcommand {\beq}{\begin{equation}}
\newcommand {\eeq}{\end{equation}}

\begin{document}


\date{\today}

\title{
Temperature-dependent magnetism in Fe foams via spin-lattice dynamics
}

\author{Robert Meyer}
\affiliation{%
	Physics Department and Research Center OPTIMAS,
	University Kaiserslautern,
	Erwin-Schr{\"o}dinger-Stra{\ss}e, D-67663 Kaiserslautern, Germany}

\author{Felipe Valencia}
\affiliation{%
  Departamento de Computaci\'on e Industrias, Facultad de Ciencias de la Ingenier\'ia, Universidad Cat\'olica del Maule, Talca.}
\affiliation{%
	Centro para el Desarrollo de la Nanociencia y la Nanotecnolog?\'ia, CEDENNA, Avda. Ecuador 3493, Santiago, Chile 9170124}

\author{Gonzalo dos Santos}
\affiliation{CONICET and Facultad de Ingenier\'ia, Universidad de Mendoza, Mendoza, 5500 Argentina}

\author{Romina Aparicio}
\affiliation{CONICET and Facultad de Ingenier\'ia, Universidad de Mendoza, Mendoza, 5500 Argentina}

\author{Eduardo M. Bringa}
\affiliation{CONICET and Facultad de Ingenier\'ia, Universidad de Mendoza, Mendoza, 5500 Argentina}
\affiliation{Centro de Nanotecnolog\'ia Aplicada, Facultad de Ciencias, Universidad Mayor, Santiago, Chile 8580745}

\author{Herbert M. Urbassek}
\email{urbassek@rhrk.uni-kl.de}
\homepage{ http://www.physik.uni-kl.de/urbassek/}
 \affiliation{%
Physics Department and Research Center OPTIMAS,
University Kaiserslautern,
Erwin-Schr{\"o}dinger-Stra{\ss}e, D-67663 Kaiserslautern, Germany}

\begin{abstract}
Spin-lattice dynamics is used to study the magnetic properties of Fe foams. The temperature dependence of the magnetization in foams is determined as a function of the fraction of surface atoms in foams, $\ns$. The Curie temperature of foams decreases approximately linearly with $\ns$, while the critical exponent of the magnetization increases considerably more strongly. If the data are plotted as a function of the fraction of surface atoms, reasonable  agreement with recent data on vacancy-filled Fe crystals and novel data on void-filled crystals is observed for the critical temperature. Critical temperature and  critical exponent also depend on the coordination of surface atoms. Although the decrease we find is relatively small, it hints to the possibility of improved usage of topology to taylor magnetic properties.

\end{abstract}

\maketitle

\section{\label{s_intro} Introduction}
Recent developments in the technique of coupled classical atomistic and spin dynamics simulations \cite{AKH*96,MDW16,MHD*19,Eva20}, also called Spin-Lattice Dynamics (SLD) simulations, have enabled calculations of the magnetization dynamics of large-scale metallic systems. While the magnetic properties of bulk metallic systems have long been studied and understood \cite{PA01}, the magnetic properties of defective and nanostructured metallic systems, such as metal foams, have not been studied in depth so far.

The effect of  isolated point defects on magnetic properties of bulk materials can be studied by ab-initio techniques \cite{SYMW00,WSNB00,DB01,FWO04,NHD06}. Similarly, the magnetic characteristics of planar surfaces have been studied \cite{ASMJ94} since they need only small simulation cells. This is not possible, however, for extended defects or for systems containing a large fraction of internal or external surfaces, since the computational effort would be too demanding. Here SLD simulations may prove useful. This technique has been used to study magnetism in Fe nanoparticles, the effect of vacancies on spin waves, the influence of magnetism on phase transitions, and other phenomena \cite{SOI*15,MDW17,MEP*17,HTM*19,SAL*20,NTAL21,SMA*21,NWC*21}.

Metal foams provide interesting possibilities as a material due to their lightweight build and thermal properties \cite{CJL*18}. Compared to other materials such as aluminum, iron foams exhibit higher strength and lower costs \cite{MAK14}.  Recently, ferritic steel foams with microscale porosity have also been studied \cite{KMK*22}. In addition, iron foams will also exhibit magnetic properties, since Fe is  ferromagnetic. However, the detailed magnetic characteristics, especially in regard to its temperature dependence, appear to have only been the subject of a small amount of research.

In this paper, we study the temperature dependent magnetization of Fe foams with varying fractions of surface atoms  and demonstrate their influence on the Curie temperature and the critical exponent. In addition, we show that the Curie temperature and the critical exponent exhibit a linear correlation with the surface atom density. In comparison with previous results obtained for isolated vacancies, we also demonstrate a difference in the critical magnetic behavior depending on the topology of the defects present in the material.

\section{\label{s_meth} Methods}

\subsection{Construction of the foams and other defective structures}

\begin{table}
	\begin{center}
		\caption{
			Parameter $H$, $M$ and $\epsilon$  used in the construction of the foams (see text), as well as porosity $p$ and resulting fraction of surface atoms, $\ns$, \qeq{eq:nsurf}.
		}
		\label{T_FV}
		\begin{tabular}{|r|r|r|r|r|}\hline
			$H^2$ & $M$ & $\epsilon$& $p$ & $\ns$ (\%) \\ \hline
			3 & 3& $-0.1$ & 0.45 & 8.0  \\
			4 & 5 & 0.0 & 0.50 & 12.3  \\ 
			9 & 17 & 0.0 & 0.50 & 13.7 \\
			11 & 12& 0.0 & 0.50 & 15.4  \\
			14 & 24& 0.0 & 0.50 & 17.2  \\
			16 & 5 & 0.0 & 0.50 & 12.7  \\
			17 & 28& 0.0 & 0.50 & 19.8 \\
			21 & 24& 0.0 & 0.50 & 19.5 \\
			30 & 24& 0.0 & 0.50 & 20.5  \\ 
			30 & 24& 0.0 & 0.50 & 21.5  \\ \hline

		\end{tabular}
	\end{center}
\end{table}

\begin{figure*}
	\begin{center}
		\includegraphics[width=0.45\linewidth]{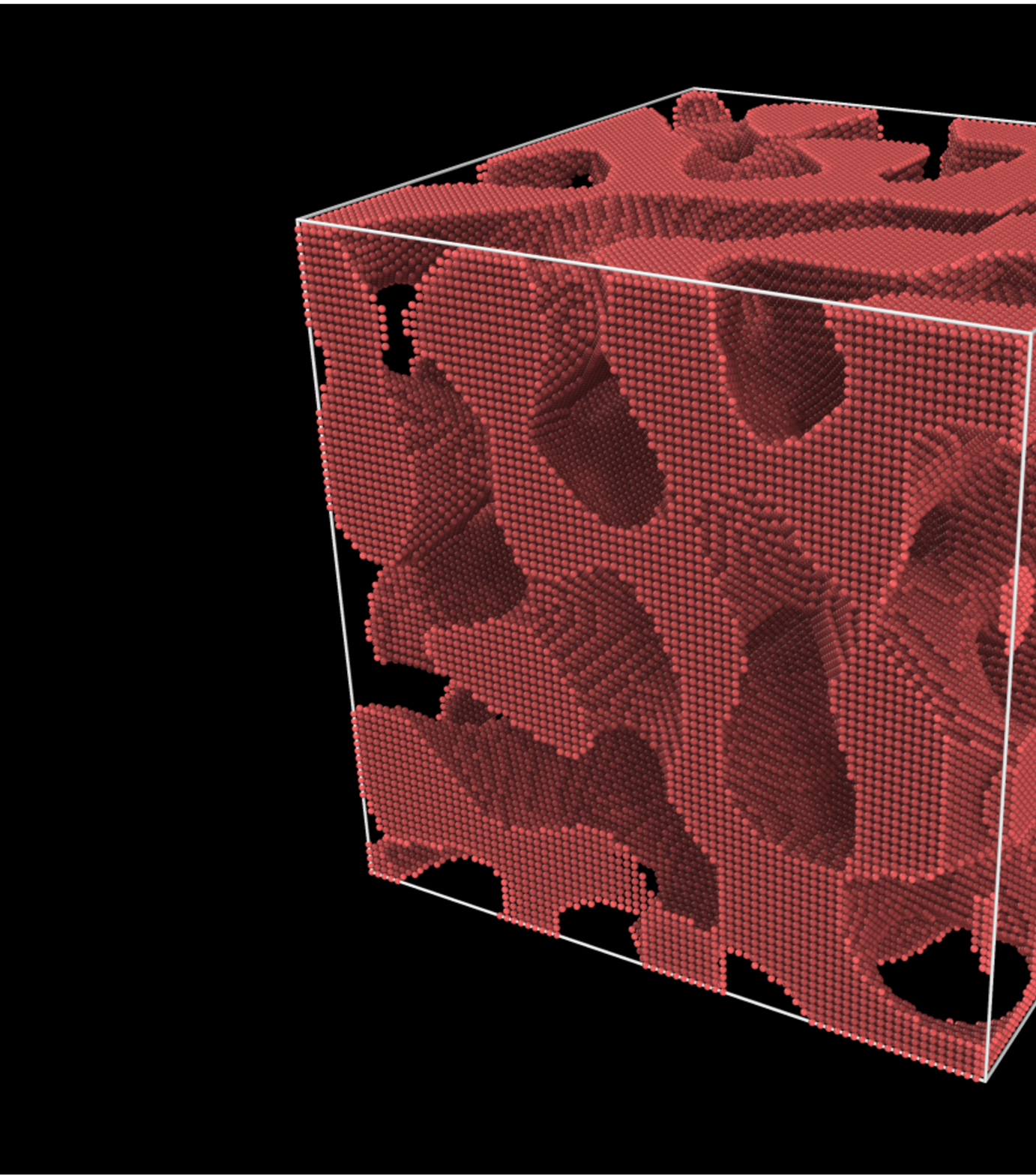}
		\includegraphics[width=0.45\linewidth]{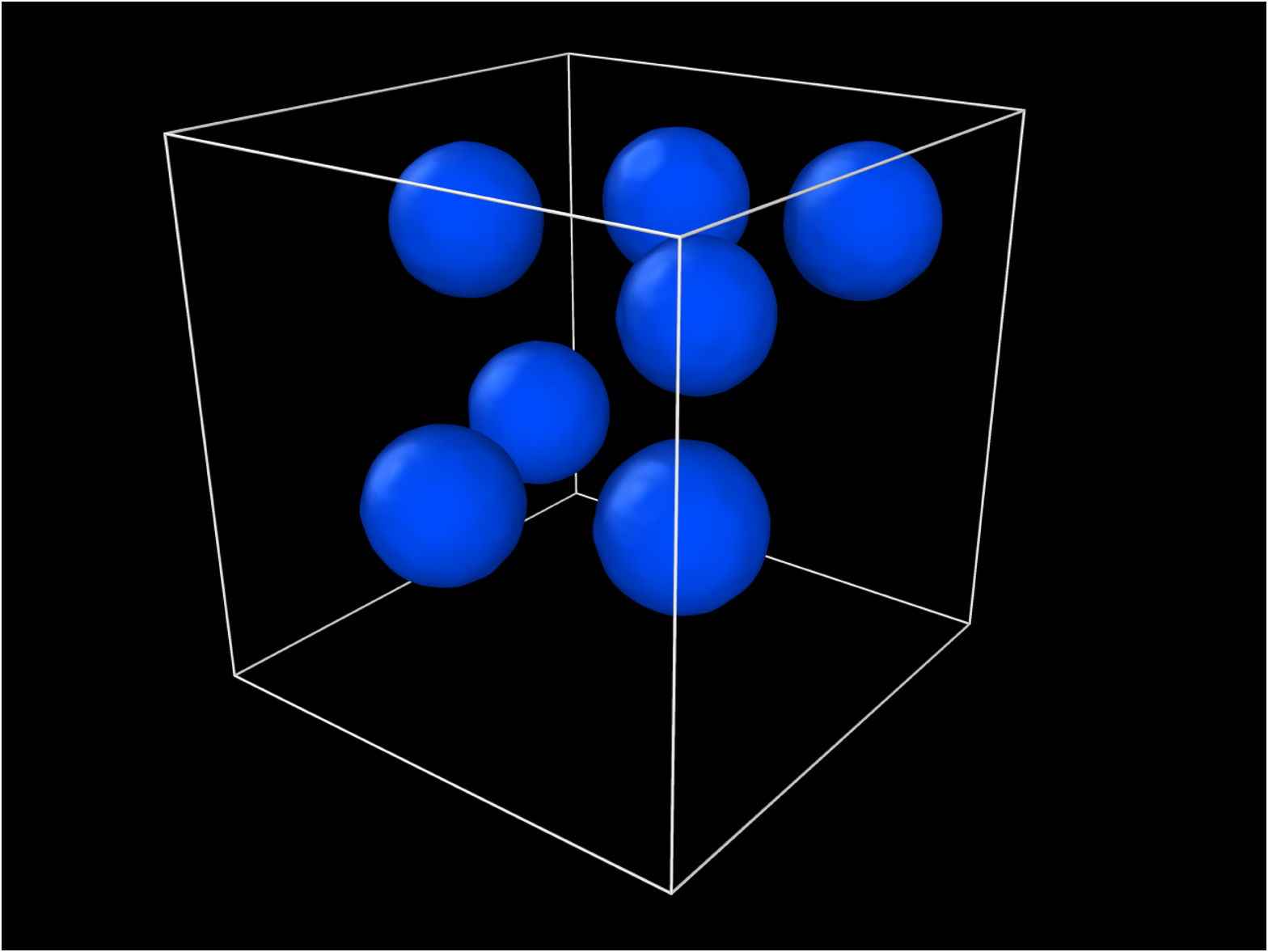}
		\caption{Representative examples of defective Fe crystals used in this study. Left: bicontinuous Fe foam with $\ns=17.2\%$. Right: Fe crystal containing 7 voids with a diameter of 6 nm and $\ns=1.9$ \%. Only the surface atoms of the voids are shown for clarity (blue color).
		}
		\label{f1}
	\end{center}
\end{figure*}

We construct Fe nanofoams using the method described by Soyarslan \etal \cite{SBPW18}. This algorithm uses the spinodal decomposition of a binary alloy as a model for the bicontinuous microstructure of a nanofoam employing a superposition of periodic waves with random directions,
\beql{e_f}
f (\bm{r})=\sqrt{\frac{2}{M}}\sum_{i=1}^M\cos(\bm{q_i}\cdot \bm{r}+\phi_i) ,
\eeq
where $\bm{r}$ represents the position vector, $M$ is the number of waves used and 
 $\phi_i$ and $\bm{q}_i$ denote the  phase and direction of wave $i$, respectively.
 In Soyarslan's algorithm \cite{SBPW18}, the  wave vectors   are obtained from the rule
\beq
 |\bm{q}|=\left|\frac{2\pi}{L}(h,k,l)\right|=\frac{2\pi}{L}H 
\eeq
with a fixed parameter $H=\sqrt{h^2+k^2+l^2}$, while $\phi_i$ are taken as random. The integers $h$, $k$, and $l$  represent  Miller indices and $L$ is the simulation cell size. This construction makes the function $f (\bm{r})$ periodic with length $L$, and the parameter $H$  defines the number of periodic waves included in $f(\bm{r})$. 
In our simulations, $L=22.8$ nm (equivalent to $80a_0$ where $a_0$ is the lattice constant of Fe) and $H$
is varied in order to produce nanofoams with different microstructures, see \qtable{T_FV}. 

Foams are characterized by their porosity $p$ defined as the ratio of the void volume to the total volume of the foam, and by  the fraction of surface atoms, $N_{\mathrm{surf}}$, relative to the  total number of atoms, $N_{\mathrm{total}}$, 
\beql{eq:nsurf} \ns=\frac{N_{\mathrm{surf}}} {N_{\rm{total}}}\, .
\eeql
The foam porosity $p$ is controlled by the function $f$, \qeq{e_f} in the following way:  a bcc lattice is set up within the box where $f$ is defined, and only the atoms whose positions $\bm{r}$ satisfy $f(\bm{r})>\epsilon$, with $\epsilon$ a control parameter, belong to the nanofoam; otherwise the atom position is left vacant. 
Thus, setting $\epsilon=0$ gives nanofoams with $p=0.5$. In the present study, we keep the nanofoam porosity  close to  $p=0.5$; different foams vary by their surface atom fraction $n_s$, which is controlled by the parameter $H$. The nanofoam specimens used in our study are assembled in \qtable{T_FV}.

For comparison of the magnetic properties in Fe foam with other defective Fe structures, we also build Fe samples containing voids. Here, the main characteristics  is the porosity; values  of $p=0.028$,  0.065 and 0.085 are realized.

For the construction of void-filled samples,  a cubic bcc Fe sample with periodic boundary conditions in all directions and volume $V=(80a_0)^3$ is set up. Spherical voids are generated at random positions by  deleting the atoms in the selected regions. For a given sample, the void diameter $D$ is kept constant and the minimum separation between voids is $D + 4a_0$ in order to prevent the voids  from coalescing after thermalization. Porosity is varied by changing the number of voids and their diameter. The sample with $p=0.028$ contains 10 voids of diameter $D=4$ nm, the one with $p=0.065$ has 7 voids with $D=6$ nm and the one with $p=0.085$, 9 voids with $D=6$ nm. We did not produce void-filled samples with higher porosity as these tended to be unstable, showing void collapse, at higher temperatures.

We verified that both the Fe foam and the void structures studied here remain thermally stable up to temperatures of 1100 K for times of  at least 50 ps. For these samples, the number of surface atoms is determined with the atomistic analysis tool OVITO \cite{Stu14}. 

We also compare our results on foam structures with the magnetic properties of bulk Fe samples filled with a predefined fraction of vacancies, $n_v$ \cite{MSA*22}. Since each vacancy  is surrounded by 8 Fe atoms, the surface atom fraction of vacancy-filled samples is determined as 
\beql{e_vac} \ns =8 n_v \,.
\eeq

After construction,  the samples are relaxed for a time period of 50 ps at $T=300$ K with pressure of 0 bar, using a time step of 1 fs. The temperature control is achieved using a Langevin thermostat, while the pressure is controlled using a Berendsen barostat.

\subsection{Spin-lattice dynamics}

SLD \cite{Eva20,SMA*21} is based on the idea that classical molecular dynamics is used to describe the  thermal  motion of atoms while the dynamics of the spin system is based on the stochastic Landau-Lifshitz-Gilbert equation. Both systems are coupled by the exchange interaction.

In detail, our SLD approach uses a magnetic spin Hamiltonian combined with the atomistic EAM potential by Chamati \etal \cite{CPMP06}. This approach is  based on the methodology presented in Ref.\  \cite{SMA*21}. The spin per atom (magnetization) of the sample is  defined as
\beql{e_S} S= \frac 1N \left| \sum_i {\bm s}_i \right| , \eeql
with $\bm{s}_i$ as the unit spin vector of atom $i$ and $N$ the total number of atomistic spins in the system. As a result, the spin per atom is $S=1$ in the ferromagnetic ground state at temperature $T=0$ K.

The magnetic spin Hamiltonian includes two  interaction terms. The exchange interaction, defined as 
\beql{e_ex} H_{\rm ex}=-J(r_{ij}) ({\bm s}_i \cdot {\bm s}_j -1) \,,\eeql 
dominates the spin-spin interaction. The exchange parameter $J(r_{ij})$ depends on the spatial distance between the spins $i$ and $j$ and therefore explicitly couples the atomistic and the spin dynamics. Consequently, the atomic motion can be influenced by the spin rotation and the spin rotation can in turn be affected by atomic motion. The spatial dependence of the exchange interaction is described by a Bethe-Slater curve and fitted to data by Ma \etal \cite{MWD08}, as performed before by dos Santos \etal \cite{SMA*21} for a bulk system and by Mudrick \etal \cite{MEP*17} for a bulk system with vacancies.  The `$-1$' term in \qeq{e_ex} avoids double counting of the effective contribution of magnetism to the total energy from the EAM potential.

In addition to the exchange interaction, we also included cubic magnetic anisotropy
\beql{h_cubic}
\begin{split}
	H_{\rm cubic} = \sum_{i=1}^{N} K_1 [ \left(\bm{s}_i \cdot \bm{n}_1\right)^2 \left(\bm{s}_i \cdot \bm{n}_2\right)^2 + \left(\bm{s}_i \cdot \bm{n}_2\right)^2 \left(\bm{s}_i \cdot \bm{n}_3\right)^2 \\
	+ \left(\bm{s}_i \cdot \bm{n}_1\right)^2 \left(\bm{s}_i \cdot \bm{n}_3\right)^2 ] + K_2 \left(\bm{s}_i \cdot \bm{n}_1\right)^2 \left(\bm{s}_i \cdot \bm{n}_2\right)^2 \left(\bm{s}_i \cdot \bm{n}_3\right)^2\,,
\end{split}
\eeql
with anisotropy constants $K_1=3.5$ eV/atom and $K_2=0.36$ eV/atom \cite{CG11} where $\bm{n}_1=(100)$, $\bm{n}_2=(010)$ and $\bm{n}_3=(001)$. 

In order to analyze the temperature dependence of the spin per atom, the crystals were heated to various temperatures between 10 K and 1100 K and then equilibrated for 10 ps. The SPIN package of the LAMMPS simulation tool was used to execute all SLD simulations \cite{TPTT18,Pli95}.
\section{\label{s_res} Results}

\subsection{\label{sec:magnetization_temperature} Temperature dependence}

\begin{figure*}
	\begin{center}
		\includegraphics[width=0.99\linewidth]{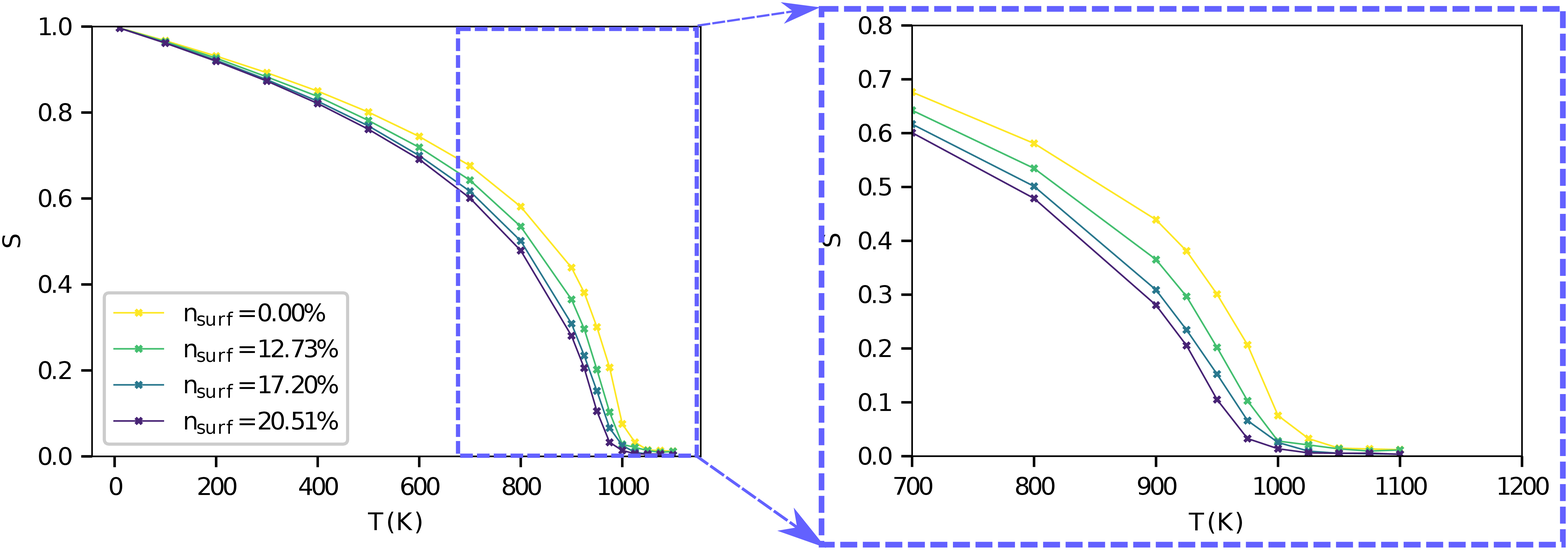}
		\caption{Temperature dependence of the spin per atom $S$ for several foams with surface atom densities $\ns$ as indicated. The inset zooms into the region around the Curie temperature. 
		}
		\label{f2}
	\end{center}
\end{figure*}

\begin{figure*}
	\begin{center}
		\includegraphics[width=0.99\linewidth]{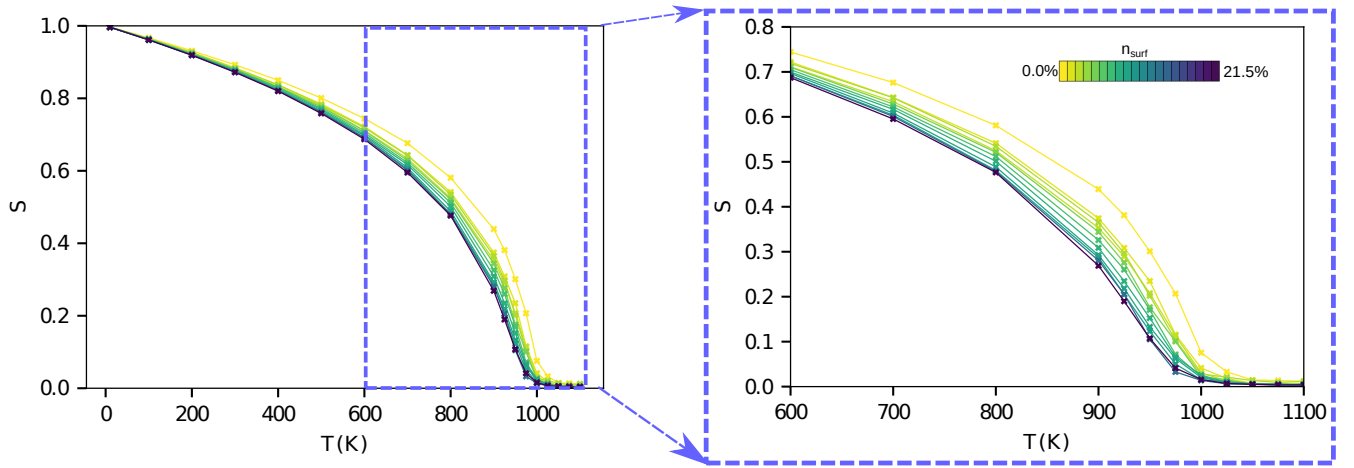}
		\caption{Temperature dependence of the spin per atom $S$ for all investigated surface atom densities, color coded according to $\ns$, see the color bar in the figure. The inset zooms into the region around the Curie temperature.
		}
		\label{f3}
	\end{center}
\end{figure*}

\begin{figure}
		\includegraphics[width=0.9\linewidth]{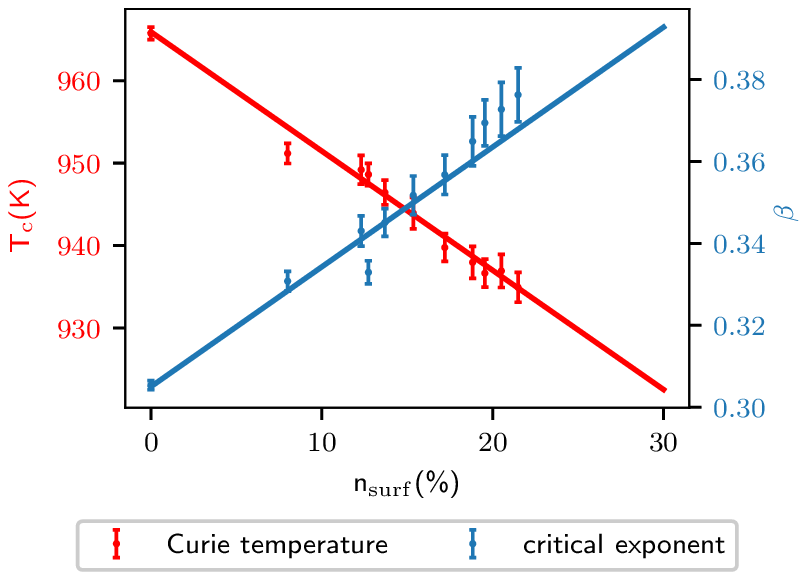}
		\caption{Dependence of the Curie temperature $T_c$ and the critical exponent $\beta$ on the surface atom fraction $\ns$ for all investigated foam structures.
		}
		\label{f4}
\end{figure}

The magnetization is determined for several different temperatures between 10 K and 1100 K, using steps of 100 K. In the critical interval between 900 K and 1100 K, we added further temperatures with  steps of 25 K. Each sample is heated to the requested temperature within 20 ps and then equilibrated  for 10 ps. This is long enough to achieve equilibrated values of the magnetization.

\qfig{f2} shows the temperature dependence of the  magnetization of the Fe foam samples for various surface atom fractions $\ns$. An increase of the fraction of surface atoms leads to a decrease of the spin per atom. This decrease is especially prominent in the high temperature region between 700 K and 1000 K. 

This trend can be even better observed in \qfig{f3}, where all resulting magnetization curves are plotted simultaneously. They are color-coded according to their surface atom fraction $\ns$ and show a continuous weakening of the spin per atom, and therefore also the magnetization, with increasing $\ns$. 

This effect is quantitatively analyzed by estimating the Curie temperature $T_c$ and the critical exponent $\beta$.
To this purpose, we fit our results for the spin per atom $S(T)$ to the polynomial decay function
\beql{s_fit} S(T)=\left( 1-\frac{T}{T_c}\right)^{\beta} \, . \eeql
In order to apply the fit function only to the ferromagnetic temperature domain, we fit the data only in the temperature range of 0--950 K. Above this temperature, paramagnetic behavior takes over and the polynomial fit \qeq{s_fit}, is not valid.

The dependence of the resulting fit values on the surface atom fraction $\ns$ is presented in \qfig{f4}. We  observe an approximately  linear relationship between the critical parameters $T_c$ and  $\beta$ on  $\ns$. While the Curie temperature is lowered slightly with increasing $\ns$, the critical exponent rises significantly to up to 1.4 times the predicted literature value for bulk Fe \cite{PV02}, $\beta=0.33$. These results are qualitatively in line with previous results for defective Fe samples with vacancies by Meyer \etal \cite{MSA*22}. In the following, a quantitative comparison between the two defect types is performed.

\subsection{\label{sec:topology_comparison} Comparison to data for void- and vacancy-filled samples}

\begin{figure*}
	\begin{center}
		\subfigure[]{\includegraphics[width=0.49\linewidth]{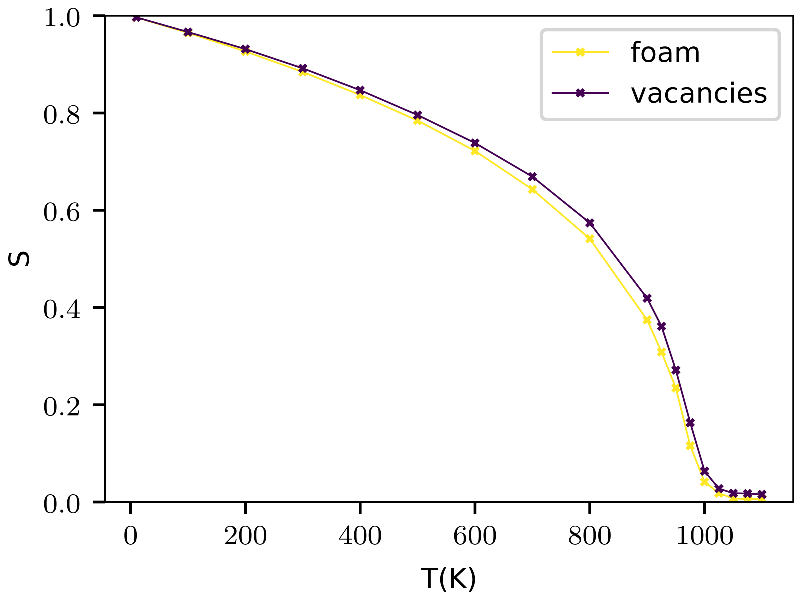}}
		\subfigure[]{\includegraphics[width=0.49\linewidth]{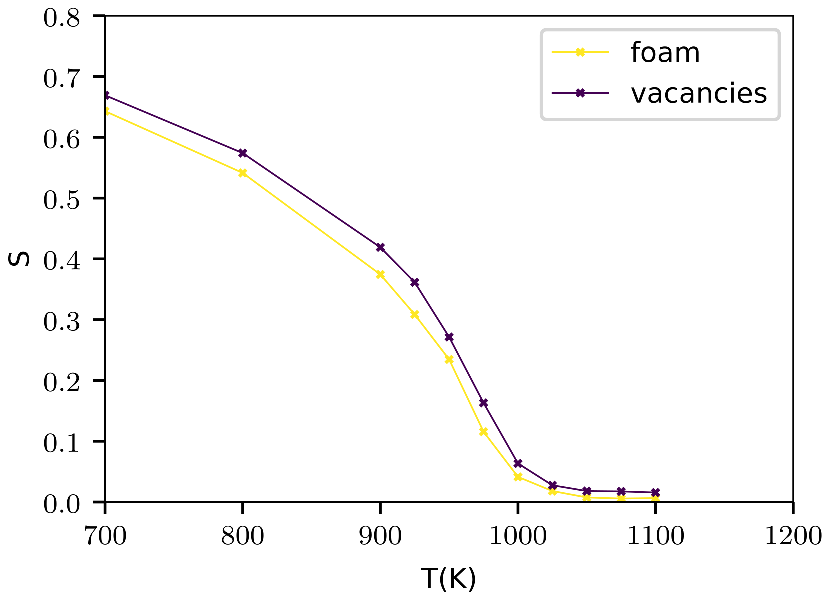}}
		\caption{
			(a) Comparison of the temperature dependence of a foam and a vacancy-filled crystal with $\ns=8.0\%$ each. Symbols denote simulation results and the line is the fit, \qeq{s_fit}. Subpanel (b) zooms into the high-temperature regime.
		}
		\label{f78}
	\end{center}
\end{figure*}

\begin{figure*}
	\begin{center}
		\subfigure[]{\includegraphics[width=0.49\linewidth]{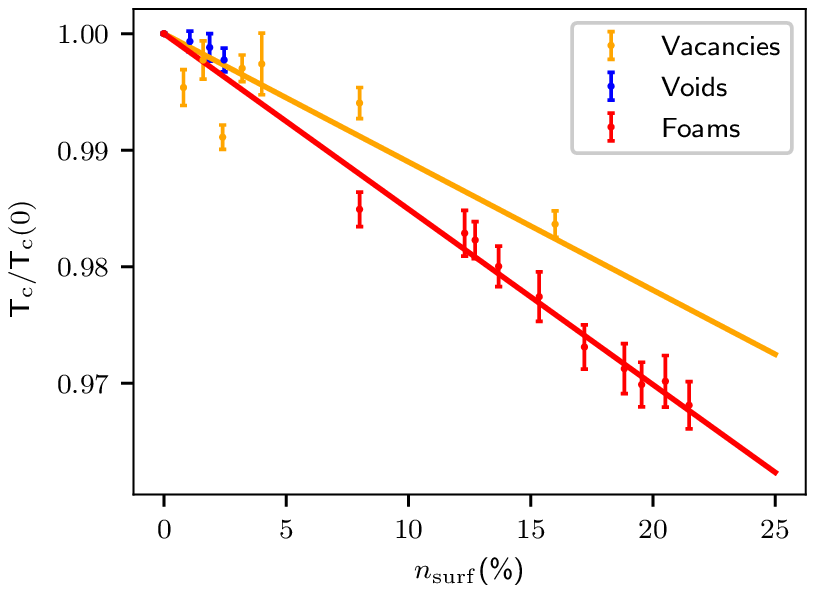}}
		\subfigure[]{\includegraphics[width=0.49\linewidth]{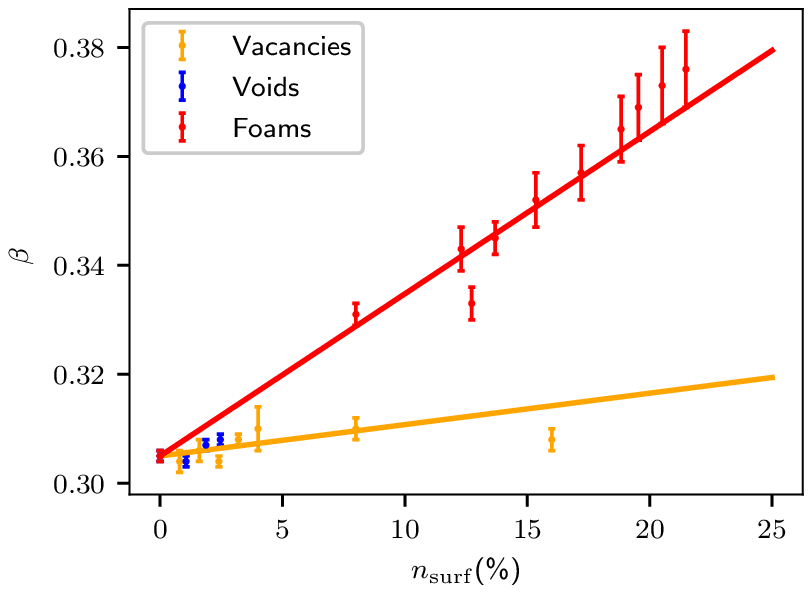}}
		\caption{Correlation between (a) the Curie temperature $T_c$ and (b) the critical exponent $\beta$ with the surface atom fraction $\ns$ for three different defect types. Lines represent least-square fits to the simulation results.
		}
		\label{f56}
	\end{center}
\end{figure*}

The foams  considered in this study have both contingent void volumes and contingent metal volumes; they are termed `bicontinuous'. In contrast, the void filled samples and also the vacancy-filled crystals have a contingent metal phase, while the unoccupied lattice sites fill unconnected disjoint void volumes. We study to what extent the magnetic properties depend on this topological difference. For further use, we denote the bicontinuous foam structures as `open' topology, and the vacancies and voids as defects with a `closed' topology.

\qfig{f78} directly compares the simulation results for a foam and a vacancy-filled crystal for the same fraction of surface atoms, $\ns=8.0$ \%. Even though the surface fraction is identical, the temperature dependence of the magnetization is different for the two systems.
This plot also demonstrates that the fit function used, \qeq{s_fit}, is appropriate, with the exception of the near vicinity of  the Curie temperature.

In   \qfig{f56}, the critical parameters,  as obtained for foams and void-filled samples are compared with the data obtained for vacancy-filled samples \cite{MSA*22}  are plotted against $\ns$. While the foam data are quite well aligned on approximately linear dependencies on the surface atom fraction, $\ns$, the data for vacancy-filled samples show a larger scatter around the linear dependence. However, a discrepancy between vacancy-filled samples on the one hand and foams of the other hand are apparent, in particular for the critical exponent, $\beta$; this discrepancy points at an influence of topology on the magnetization results. In addition, we note  that the data for void-filled samples are well aligned with the data for vacancy-filled samples. The surface atom fraction of void-filled samples are quite small, since the voids contain a large number of unoccupied atom sites in their interior. However, the good alignment of the data points for void- and vacancy-filled samples is reassuring as it shows that our determination of the surface atom fraction for  vacancy-filled samples, \qeq{e_vac}, is appropriate.

We quantify the correlation between the critical parameters and the surface atom fraction $\ns$ by using the simple linear relationships
\begin{subequations} \begin{align}
	T_c  & = T_c(0)\, (1 - \alpha_{T} \ns) \, ,  \\
	\beta & = \beta(0) (1 + \alpha_{\beta} \ns) \, , 
\end{align} \label{eq:lin_critical} \end{subequations}
where $T_c(0)$ and $\beta(0)$ denote the values of $T_c$ and $\beta$, respectively, for a defect-free Fe crystal.
The results of the fitting procedure are summarized in \qtable{tab:crit_param}. The correlation coefficient $\alpha_{T}$ is, within the margin of error, similar for open and closed topologies; in particular the large scatter of the data for a vacancy-filled sample does not allow to determine a clear difference.

\begin{table}
	\begin{center}	
		\caption{Correlation parameter $\alpha_i$ according to \qeqs{eq:lin_critical}, as presented in \qfig{f56}.}
		\label{tab:crit_param}
		\begin{tabular}{|c||c|c|}
			\hline
			system  & $\alpha_T$ &  $\alpha_{\beta}$ \\ \hline
			
			foams &  0.151 $\pm$ 0.004 &  0.976 $\pm$ 0.041 \\
			
			vacancies &  0.110 $\pm$ 0.042 &  0.189 $\pm$ 0.065 \\
			\hline
		\end{tabular}
	\end{center}
\end{table}

The second correlation coefficient, $\alpha_{\beta}$, however,  differs greatly for open and closed topologies, the difference amounting to a factor of around 5. This is also illustrated in \qfig{f56}, where we can observe a largely different trend for the critical exponent of open and closed defect topologies, but only a slight difference for the Curie temperature. This difference shows up in the magnetization curves, \qfig{f78}, in the vicinity of the critical point: The smaller exponent for the vacancy-filled samples features a more abrupt decay of the magnetization near $T_c$.

The size of $\alpha_T$ can be rationalized by assuming a simple  interpolation of the Curie temperature for bulk atoms, $T_c(0)$, and the one for surface atoms. 
Assuming that surface atoms are characterized by a Curie temperature that is smaller by a factor $\gamma$ than the bulk Curie temperature, such a linear interpolation yields
\begin{align} T_c & = (1- \ns) T_c(0) + \ns \, \gamma T_c(0 )\nonumber \\
& = [1-(1-\gamma) \ns ] \, T_c(0),  \label{e_T} \end{align} 
and hence $\alpha_{T} = 1-\gamma$. If an ensemble of surface atoms were nonmagnetic, $\gamma=0$, the decline in $T_c$ in porous materials caused by surface atoms would be strongest, $\alpha_{T} =1$. Previous simulations of magnetic nanoparticles \cite{SAL*20} suggest a value of $\gamma=0.5$, resulting in $\alpha_{T} = 0.5$. Our present results indicate that an ensemble of surface atoms has even a higher $T_c$ characterized by $\gamma= 0.8$--0.9 and hence $\alpha_{T} = 0.1$--0.2.

A similar argument for $\alpha_{\beta}$ does not appear to hold true. This is plausible since $\beta$ is an exponent determining the shape of the magnetization curve, which will be more susceptible to details of the modeling, while $T_c$ only sets the temperature scale of the magnetization curve.

To help explain the differences in \qfig{f56}, one can consider how topology affects number of neighbors. `Surface atoms' associated with vacancies will have ($Z-1$) neighbors, where $Z=8$ is the bulk coordination. On the other hand, atoms on a flat (001) surface will have $Z/2$ neighbors for a cubic system. 
Voids will have curvature in 3 dimensions, leading to some intermediate value of coordination, close to vacancies and around 6 for the cases we studied. Foams could be considered interconnected nanowires, with several different surface orientations. In our case the average coordination of foam surface atoms is around 5.5, lower than for voids (5.9).
In the simple Ising model critical temperature is proportional to coordination, so foams would have lower critical temperature than voids or vacancies. The critical exponent is inversely proportional to dimensionality in the Ising model, which could be associated with connectivity of the spins, and would be higher for vacancies than for foams, explaining the trend in our results.

In general, topology affects materials properties \cite{GS14a,GS18}, and here we show that, although trends are similar, curvature gives different scaling laws between porosity by vacancies or voids (with positive curvature) and porosity in bicontinuous nanostructures, with negative Gaussian curvature \cite{LB19}.

\section{\label{s_conc} Conclusions}

In this paper, we used SLD to study the temperature dependence of the magnetism in Fe foams. 
 Fe foams show a slight decrease  of the Curie temperature and a significant increase of the critical exponent $\beta$ with increasing surface atom fraction,  $\ns$.
These findings are in qualitative agreement with previous results for vacancy-filled Fe crystals \cite{MSA*22} and with novel data on void-filled crystals. The agreement is even quantitatively satisfactory for the Curie temperature. However, the critical exponent $\beta$ shows a fourfold stronger dependence on the surface atom fraction,  $\ns$, for foams than for vacancies or voids.

It may be speculated that the origin of the different dependencies of the critical exponent $\beta$ for foams and vacancy- or void-filled crystals originates in their different topologies. While our foams a bi-continuous -- that is both the material and the vacuum phase are contiguous throughout the sample -- the vacuum phase in the vacancy- and void-filled crystals are not contiguous. This might be at the origin of the different behavior in $\beta$ seen in this study, together with the fact that the connectivity of surface atoms is significantly different in bicontinuous foams than for isolated voids or vacancies. 

This study assumes the same magnetic moment and exchange function for surface atoms. This is only an approximation. Recently, variations of the magnetic moment with effective electronic density were studied for the surface of Fe nanoparticles \cite{SMA*21}. Metallic nanofoams might suffer oxidation \cite{GBR18}. Fe surfaces might form magnetic oxides or form a magnetically dead thin layer \cite{UUS*17,PSFF86}. 
As a first rough approximation, this could be modeled as a shell of non-magnetic Fe atoms, which would also decrease Curie temperature.

This study focus on a model foam, but alternative bicontinuous foams could be considered in the future \cite{BKM*18}, including magnetic foams where the pores are replaced by non-magnetic material, similar to CuMo foams \cite{BCFM19}, or when two magnetic materials are used, as in FeCr foams \cite{XLSJ20,MBD*20}. 

Results presented here show that nanoscale pore topology could help tayloring magnetic properties, alongside other variables already in use, like defect \cite{LDB21} or impurity content \cite{TFS20} or microstructure \cite{QGY*21}.

\begin{acknowledgments}
Funded by the Deutsche Forschungsgemeinschaft (DFG, German Research Foundation) --  project number 268565370 -- TRR 173  Spin+X (project A06).
Simulations were performed at the High Performance Cluster Elwetritsch (Regionales Hochschulrechenzentrum, TU Kaiserslautern, Germany). 
EMB thanks support from ANPCyT PICTO-UUMM-2019-00048 and SIIP-UNCUYO 06/M104 grants.
FV thanks Fondo Nacional de Investigaciones Cient?\'ificas y Tecnol\'ogicas (FONDECYT, Chile) under grants \#1190662 and  \#11190484. FV also  thanks Financiamiento Basal para Centros Cient??ficos y Tecnol\'ogicos de Excelencia AFB180001. Powered@NLHPC: this research was partially supported by the supercomputing infrastructure of the NLHPC (ECM-02)

\section*{Data availability}
This study is based only on data obtained using the methods described in this paper.

\end{acknowledgments}

\bibliography{bibo/string,bibo/all,bibo/publ}

\clearpage\newpage

\end{document}